\documentclass[%
amsmath,amssymb,
aps, physrev,
prl
floatfix,
twocolumn
]{revtex4-2}

\usepackage{graphicx}
\usepackage{dcolumn}
\usepackage{bm}
\usepackage{hyperref}
\usepackage[mathlines]{lineno}
\usepackage{appendix}
\usepackage{comment}
\usepackage{xcolor}

\begin{document}



\title{Equal-spin and opposite-spin density-density correlations in the BCS-BEC crossover: 
Gauge Symmetry, Pauli Exclusion Principle, Wick's Theorem and Experiments}

\author{Nikolai Kaschewski}
\affiliation{Physics Department and Research Center OPTIMAS, Rhineland-Palatinate Technical University Kaiserslautern-Landau, 67663 Kaiserslautern, Germany}

\author{Axel Pelster}
\affiliation{Physics Department and Research Center OPTIMAS, Rhineland-Palatinate Technical University Kaiserslautern-Landau, 67663 Kaiserslautern, Germany}

\author{Carlos A. R. {S\'a} de Melo}
\affiliation{School of Physics, Georgia Institute of Technology, Atlanta, 30332, USA}

\date{\today}

\begin{abstract}
We develop a general theory of 
{\it spin}-dependent density-density correlations, 
that is valid for any temperature, interactions, dimensions and mass or population status
of Fermi gases with two internal states. 
We use gauge invariance and the Pauli principle to 
establish constraints on the {\it spin}-dependent density-density correlations that are consistent with the fluctuation-dissipation and Wick's theorem.
As an example, we study
the {\it spin}-dependent density-density correlations from the BCS to the Bose regime in two dimensions at zero temperature, inspired by experiments in 
$^6{\rm Li}$. We show that two-particle irreducible contributions, involving collective excitations, many-particle scattering and vertex corrections, are essential to describe experiments. In particular, 
the two-particle irreducible terms are responsible for the emergence of an experimentally observed minimum in the opposite-{\it spin} density-density correlations.
\end{abstract}

\maketitle


\textit{Introduction:} Ultracold Fermi gases represent a versatile platform to investigate fundamental properties of quantum many-body physics,
such as correlations 
that have been explored in condensed matter~\cite{Kuroki-2024,Cazalilla-2024,Nygard-2024,Bouzerar-2025,Yelin-2025},
nuclear physics~\cite{Isayev-2008,Berger-2010,Schuck-2011,Sandulescu-2018,Schuck-2020}, 
astrophysics~\cite{Ramos-2005,Liu-2024,Sedrakian-2024,Chamel-2025}, and ultracold gases~\cite{Hulet-2004,Giorgini-2004,Stringari-2006,Carlson-2011,Castin-2012,He-2016,Romero-Rochin-2022,Romero-Rochin-2023}, where the latter serve as quantum simulators~\cite{Bloch-2012,Bloch-2017,Ohashi-2018,Zoller-2022}. The measurement of correlations
provide deep insight into degenerate
Bose~\cite{Kheruntsyan-2012,Schmiedmeyer-2017,Berges-2020,Schmiedmayer-2021} and Fermi~\cite{Jin-2005,Bloch-2005,Ketterle-2008,Moritz-2018,Moritz-2022,Pan-2024} systems. 
Due to existence of Fano-Feshbach resonances~\cite{Feshbach-1958,Fano-1961,Tiesinga-2010}, a prime example of the manifestation of correlations is the BCS-BEC crossover explored both
theoretically~\cite{Engelbrecht-1993,SadeMelo-1997,Strinati-2004,Zwerger-2017,Chen-2024,Pawlowski-2024} 
and experimentally~\cite{Zwierlein-2019,Widera-2020,Widera-2023-2,Denschlag-2024,Thompson-2025,Pan-2025}.
Correlations were further investigated
in momentum-frequency space 
via collective modes
in the low-energy-long-wavelength limit~\cite{SadeMelo-1997,Stringari-2006-2,Castin-2019,Enss-2019,Kurkjian-2019,Tempere-2020,Kurkjian-2021,Moritz-2022,SaDeMelo-2023,Pan-2024}. 
However, correlations in real space and time have been little explored because of limited experimental accessibility.

With the recent development of continuous quantum gas microscopes (CQGM)~\cite{Yefsah-2025,Ketterle-2025,Zwierlein-2025,Yefsah-2025-2,Yefsah-2025-3,Yefsah-2025-4}, 
it is now possible to directly measure spatially-resolved equal-time density-density correlations
in two-dimensional (2D) Fermi gases.
These new microscopes differ from previous lattice quantum gas microscopes~\cite{Greiner-2009,Naegerl-2010,Zwierlein-2015,Zwierlein-2016,Bloch-2017,Gross-2019,Bakr-2021,Ni-2021,Zoller-2022}
that were restricted to measurements of spatial correlations
at large distances only. The CQGM has better spatial resolution than the typical interparticle distance, allowing for the 
detection of anti-bunching in real space for opposite-{\it spin} 
density-density correlations of $^6{\rm Li}$ atoms~\cite{Yefsah-2025-3} in 2D.

In this letter, we develop a theory of 
{\it spin}-dependent density-density correlations, 
that is valid for any temperature, interactions, dimensions and mass or population status of Fermi gases with two internal states. 
We impose two general constraints, gauge symmetry and Pauli exclusion principle, that must be satisfied by {\it spin}-dependent density-density correlations, and 
that are consistent with the fluctuation-dissipation and Wick's theorem.
Using these conditions, we obtain an equation of state that includes two-particle reducible and irreducible contributions. To show the power of these requirements, we investigate {\it spin}-dependent density-density correlations from the BCS to the Bose regime in 2D at zero temperature, similar to recent CQGM experiments in $^6{\rm Li}$~\cite{Yefsah-2025-3}.
We demonstrate that two-particle irreducible contributions, such as collective excitations, many-particle scattering and vertex corrections,
are essential to explain experiments.
In particular, we show that two-particle irreducible terms are responsible for the observed minimum in the opposite-{\it spin} density-density correlation function below the uncorrelated value of one~\cite{Yefsah-2025-3}.
We make quantitative predictions about 
anti-correlations by evaluating the position and depth of this minimum.

\textit{Correlation functions:} We are interested in the normalized equal-time density-density correlation functions
\begin{equation}
g_{\rm ss^\prime}({\bf r}, {\bf r^\prime})
= 
\frac{\langle {\hat n}_{\rm s}({\bf r}) 
{\hat n}_{\rm s^\prime}({\bf r}^\prime) \rangle}
{\langle {\hat n}_{\rm s}({\bf r})\rangle \langle {\hat n}_{\rm s^\prime}({\bf r}^\prime )\rangle},
\end{equation}
where ${\hat n}_{\rm s} ({\bf r}) = \hat\psi^\dagger_{\rm s} ({\bf r}) \hat\psi_s ({\bf r})$ is the density operator for {\it spin} projection ${\rm s}$ , with 
$\hat\psi^\dagger_{\rm s} ({\bf r})$ and $\hat\psi_{\rm s} ({\bf r})$ being the
field operators for fermions at position ${\bf r}$
in continuum systems.
The notation $\langle {\hat A} \rangle$ denotes the ensemble 
expectation value of operator ${\hat A}$ at temperature $T$.
We use the word {\it spin} to represent either
the internal states of identical
fermions with mass $m_{\uparrow} = m_{\downarrow} = m$ or different species of fermions with $m_{\uparrow} \ne m_{\downarrow}$.
For fermions with equal or
different masses $m_\uparrow, m_\downarrow$, there are two {\it spin} projections 
${\rm s} = \{\uparrow, \downarrow\}$, such that the equal-{\it spin} (opposite-{\it spin}) correlation function corresponds to 
$\rm s = s^\prime$ $(\rm s \neq s^\prime)$.

Before explicitly calculating $g_{\rm ss^\prime}({\bf r}, {\bf r^\prime})$ for a specific Lagrangian, we highlight a few of its symmetries that are independent of dimensionality, interactions, temperature, masses and population status of {\it spin} projections. 

First, we emphasize the invariance of  
$g_{\rm ss^\prime}({\bf r}, {\bf r^\prime})$ under the local U(1) gauge transformation 
$\hat\psi_{\rm s} {(\bf r)} \to \hat\psi_{\rm s} {(\bf r)} e^{i\phi_{\rm s} ({\bf r})}$, since the local density operator 
${\hat n}_{\rm s} ({\bf r})$ is unchanged by it. This is extremely important, because even if the system spontaneously breaks a local U(1) gauge symmetry due to 
local pairing 
$\hat\psi_\uparrow {(\bf r)} \hat\psi_\downarrow {(\bf r)} \to \hat\psi_\uparrow {(\bf r)} \hat\psi_\downarrow {(\bf r)} 
e^{i\left[\phi_{\uparrow} ({\bf r}) + \phi_{\downarrow}({\bf r})\right]}$, the physical $g_{\rm ss^\prime}({\bf r}, {\bf r^\prime})$ must be gauge invariant.

Second, we investigate the Pauli-principle.
Using $\{\psi_{\rm s} ({\bf r}), \psi_{ \rm s^\prime}^\dagger ({\bf r}^\prime) \}
= \delta_{\rm s s^\prime} \delta ({\bf r} - {\bf r}^\prime)$,
we write 
$
\langle {\hat n}_{\rm s}({\bf r}) 
{\hat n}_{\rm s^\prime}({\bf r}^\prime) \rangle
= 
\langle \mathopen{:} {\hat n}_{\rm s}({\bf r}) 
{\hat n}_{\rm s^\prime}({\bf r}^\prime) \mathclose{:} \rangle
+
\delta_{\rm s s^\prime} \delta ({\bf r} - {\bf r}^\prime)
\langle {\hat n}_{\rm s} ({\bf r}) \rangle
$
in terms of the normal ordered pair correlations
$
\langle \mathopen{:} {\hat n}_{\rm s}({\bf r}) 
{\hat n}_{\rm s^\prime}({\bf r}^\prime) \mathclose{:} \rangle
= 
\langle 
\psi^{\dagger}_{\rm s} ({\bf r}) \psi^{\dagger}_{\rm s^\prime} ({\bf r}^\prime) 
\psi_{\rm s^\prime} ({\bf r}^\prime)
\psi_{\rm s} ({\bf r})
\rangle.
$
Thus, we obtain
\begin{equation}
g_{\rm ss^\prime}({\bf r}, {\bf r^\prime})
= 
g_{\rm s s^{\prime}}^{\rm reg}({\bf r}, {\bf r^\prime})
+ 
\frac{\delta_{\rm s s^\prime} \delta ({\bf r} - {\bf r}^\prime)}
{ \langle 
{\hat n}_{\rm s^\prime}({\bf r}^\prime )\rangle},
\label{eqn:g-regular-irregular-decomposition}
\end{equation}
where the regular part is 
\begin{equation}
g_{\rm s s^{\prime}}^{\rm  reg} ({\bf r}, {\bf r^\prime})
=
\frac{\langle \mathopen{:} {\hat n}_{\rm s}({\bf r}) 
{\hat n}_{\rm s^\prime}({\bf r}^\prime) \mathclose{:}\rangle}
{\langle {\hat n}_{\rm s}({\bf r})\rangle \langle {\hat n}_{\rm s^\prime}({\bf r}^\prime )\rangle}.
\label{eqn:g-regular-definition}
\end{equation}
We emphasize that the only difference between 
$g_{\rm ss^\prime}({\bf r}, {\bf r^\prime})$ and 
$g_{\rm s s^{\prime}}^{\rm reg}({\bf r}, {\bf r^\prime})$ 
in Eqs.~(\ref{eqn:g-regular-irregular-decomposition}) and~(\ref{eqn:g-regular-definition})
is the singular term that
arises for ${\rm s} = {\rm s}^\prime$ and ${\bf r} = {\bf r}^\prime$.
The Pauli principle manifests itself 
in $\langle \mathopen{:}{\hat n}_{\rm s}({\bf r}) 
{\hat n}_{\rm s}({\bf r}) \mathclose{:}\rangle = 0$ leading
to $\lim_{{\bf r}^{\prime} \to {\bf r}} g_{\rm ss}^{\rm reg} ({\bf r}, {\bf r^\prime}) = 0$,
but $\langle {\hat n}_{\rm s}({\bf r}) 
{\hat n}_{\rm s}({\bf r}) \rangle = 
\delta ({\bf 0}) \langle {\hat n}_{\rm s} ({\bf r}) \rangle$
is singular giving 
$\lim_{{\bf r}^{\prime} \to {\bf r} } g_{\rm s s} ({\bf r}, {\bf r^\prime}) = \delta ({\bf 0})/\langle {\hat n}_{\rm s} ({\bf r}) \rangle$.

Using the decomposition
$
{\hat n}_{\rm s} ({\bf r}) = 
\langle {\hat n}_{\rm s} ({\bf r}) \rangle 
+ \delta{\hat n}_{\rm s} ({\bf r}),  
$
with the ensemble average 
$\langle {\hat n}_{\rm s} ({\bf r}) \rangle$ being equal to the local density $n_{\rm s} ({\bf r})= \langle {\hat n}_{\rm s} ({\bf r}) \rangle$, gives
\begin{equation}
g_{\rm ss^\prime}({\bf r}, {\bf r^\prime})
= 
1 + 
\frac{\chi_{\rm s s^\prime} ({\bf r}, {\bf r}^\prime)}
{n_{\rm s} ({\bf r}) n_{\rm s^\prime} ({\bf r}^\prime)},
\label{eqn:dens-dens-correl-general}
\end{equation}
where $\chi_{\rm s s^\prime} ({\bf r}, {\bf r}^\prime) 
= \langle \delta{\hat n}_{\rm s}({\bf r}) 
\delta{\hat n}_{\rm s^\prime}({\bf r}^\prime) \rangle$ 
represents the connected density-density correlation
tensor. For $g_{\rm ss^\prime}({\bf r}, {\bf r^\prime}) = 1$, the particles at ${\bf r}$ and ${\bf r}^\prime$ are uncorrelated, while for $g_{\rm ss^\prime}({\bf r}, {\bf r^\prime}) < 1$, the fermions exclude (correlation holes, anti-bunching) and for $g_{\rm ss^\prime}({\bf r}, {\bf r^\prime}) > 1$, they approach each other (clustering, pairing, bunching).
Comparing Eqs.~(\ref{eqn:g-regular-irregular-decomposition}) 
and~(\ref{eqn:dens-dens-correl-general}), we write
\begin{equation}
\chi_{\rm s s^\prime} ({\bf r}, {\bf r}^\prime)
= \chi_{\rm s s^\prime}^{\rm reg} ({\bf r}, {\bf r}^\prime)  + 
\delta_{\rm s s^\prime} \delta ({\bf r} - {\bf r}^\prime)
n_{\rm s} ({\bf r}),
\label{eqn:chi-regular-irregular-decomposition}
\end{equation}
where the regular part of $\chi_{\rm s s^\prime} ({\bf r}, {\bf r}^\prime)$ is
\begin{equation}
\chi_{\rm s s^\prime}^{\rm reg} ({\bf r}, {\bf r}^\prime) 
=
\langle\mathopen{:}{\hat n}_{\rm s}({\bf r}) 
{\hat n}_{\rm s^\prime}({\bf r}^\prime)\mathclose{:}\rangle
- n_{\rm s} ({\bf r}) n_{\rm s^\prime} ({\bf r}^\prime),
\label{eqn:chi-regular-definition}
\end{equation}
leading to the manifestation of the Pauli principle as 
\begin{equation}
n_{\rm s}^2 ({\bf r})
= 
- \lim_{{\bf r}^\prime \to {\bf r}} \chi_{\rm s s}^{\rm reg} ({\bf r}, {\bf r}^\prime). 
\label{eqn:Pauli-principle-chi-regular}
\end{equation}
The relation in Eq.~(\ref{eqn:Pauli-principle-chi-regular}) corresponds to the local equation of state (EoS) that fixes the 
{\it spin}-resolved density $n_{\rm s} ({\bf r})$.
The resulting EoS is
consistent with the fluctuation-dissipation 
theorem~\cite{SaDeMelo-2011}, that is,
\begin{equation}
\chi_{\rm s s^\prime} ({\bf r}, {\bf r}^\prime) = 
-\frac{T}{V^2} 
\frac{\delta^2 \Omega [j_{\rm s}] }{\delta j_{\rm s} ({\bf r}) \delta j_{\rm s^\prime}({\bf r}^\prime)},
\label{eqn:fluctuation-dissipation-theorem}
\end{equation}
where $\Omega [j_{\rm s}]$ stands for the
thermodynamic potential including the density-fluctuation source terms $j_{\rm s} ({\bf r})$ and $j_{\rm s} ({\bf r}^\prime)$. The tensor $\chi_{\rm s s^\prime} ({\bf r}, {\bf r}^\prime)$ is sometimes called the spatially-dependent compressibility matrix~\cite{SaDeMelo-2011}.

\textit{Wick's theorem:} 
Using Wick's decomposition for the equal-time pair correlation 
function gives
\begin{eqnarray}
\langle \mathopen{:} {\hat n}_{\rm s}({\bf r}) 
{\hat n}_{\rm s^\prime}({\bf r}^\prime) \mathclose{:}\rangle
& = &
\langle \hat{n}_{\rm s}({\bf r}) \rangle \langle \hat{n}_{\rm s^\prime}({\bf r}^\prime)\rangle
- \vert G_{\rm s s^\prime} ({\bf r}, {\bf r}^\prime )\vert^2
\nonumber \\
& + &\vert F_{\rm s s^\prime} ({\bf r}, {\bf r}^\prime ) \vert^2 
+ \langle {\hat n}_{\rm s}({\bf r}) {\hat n}_{\rm s^\prime}({\bf r}^\prime) \rangle_{\rm irr}.
\label{eqn:pair-correlation-function-Wick-decomposition}
\end{eqnarray}
Here, the first term $\langle \hat{n}_{\rm s}({\bf r}) \rangle \langle \hat{n}_{\rm s^\prime}({\bf r}^\prime)\rangle$ is the product of local densities 
$n_{\rm s}({\bf r}) n_{\rm s^\prime} ({\bf r}^\prime)$.
The second term 
represents the negative of the squared modulus of the normal (standard) Green's functions
$G_{\rm s s^\prime} ({\bf r}, {\bf r}^\prime) 
= \langle \hat{\psi}^\dagger_{\rm s}({\bf r})\hat{\psi}_{\rm s^\prime}({\bf r}^\prime) \rangle$.
The third term 
denotes the squared modulus of anomalous (pair) Green's functions $F_{\rm s s^\prime} ({\bf r}, {\bf r}^\prime ) =
\langle \hat{\psi}_{\rm s}({\bf r})\hat{\psi}_{\rm s^\prime}({\bf r}^\prime)\rangle$. 
The last term $\langle {\hat n}_{\rm s}({\bf r}) {\hat n}_{\rm s^\prime}({\bf r}^\prime) \rangle_{\rm irr} = 
\langle \delta {\hat n}_{\rm s}({\bf r}) \delta {\hat n}_{\rm s^\prime}({\bf r}^\prime) \rangle_{\rm irr}$ stands for the two-particle irreducible part of the {\it spin}-dependent density-density correlations.

Using Wick's theorem in Eq.~(\ref{eqn:pair-correlation-function-Wick-decomposition}) and the relation in Eq.~(\ref{eqn:chi-regular-definition}), we write 
\begin{equation}
\chi_{\rm s s^\prime}^{\rm reg} ({\bf r}, {\bf r}^\prime) 
= 
\chi_{{\rm red},\rm s s^\prime} ({\bf r}, {\bf r}^\prime) 
+
\chi_{{\rm irr},\rm s s^\prime} ({\bf r}, {\bf r}^\prime),
\label{eqn:red-irr-decomposition}
\end{equation}
where the two-particle reducible contribution takes the form 
\begin{equation}
\chi_{{\rm red},\rm s s^\prime} ({\bf r}, {\bf r}^\prime) 
=
- \vert G_{\rm s s^\prime} ({\bf r}, {\bf r}^\prime ) \vert^2 
+ \vert F_{\rm s s^\prime} ({\bf r}, {\bf r}^\prime ) \vert^2,
\label{eqn:reducible-chi}
\end{equation}
while the two-particle irreducible contribution reads
\begin{equation}
\chi_{{\rm irr},\rm s s^\prime} ({\bf r}, {\bf r}^\prime) 
=
\langle \delta {\hat n}_{\rm s}({\bf r}) \delta {\hat n}_{\rm s^\prime}({\bf r}^\prime) \rangle_{\rm irr}.
\label{eqn:irreducible-chi-real-space}
\end{equation}
Combining Eqs.~(\ref{eqn:g-regular-irregular-decomposition}) and~(\ref{eqn:red-irr-decomposition}), we write 
\begin{equation}
g_{\rm s s^{\prime}}^{\rm  reg} ({\bf r}, {\bf r^\prime})
= 
1 
+ 
\delta g_{\rm red, s s^{\prime}}({\bf r}, {\bf r^\prime})
+
\delta g_{\rm irr, s s^{\prime}}({\bf r}, {\bf r^\prime}),
\label{eqn:g-regular-reducible-irreducible-decomposition}
\end{equation}
with the contributions from correlations being 
$
\delta g_{\alpha,\rm s s^{\prime}}({\bf r}, {\bf r^\prime}) 
= 
\chi_{\alpha,\rm s s^\prime} ({\bf r}, {\bf r}^\prime)
/n_{\rm s} ({\bf r})  n_{\rm s^\prime} ({\bf r}^{\prime}),
$
using $\alpha \in \{ {\rm red, irr} \}$,
corresponding to deviations from one, that is the uncorrelated limit.
Furthermore, the Pauli principle from Eq.~(\ref{eqn:Pauli-principle-chi-regular}), separates the local EoS into reducible and irreducible contributions
\begin{equation}
n_{\rm s}^2 ({\bf r}) = 
-\lim_{{\bf r}^\prime \to {\bf r}}
\left[
\chi_{\rm red, ss} ({\bf r}, {\bf r}^\prime )
+
\chi_{\rm irr, ss} ({\bf r}, {\bf r}^\prime )
\right].
\label{eqn:pauli-principle-condition-local}
\end{equation}

For translationally invariant systems 
$n_{\rm s} ({\bf r}) = n_{\rm s}$, the general relations
from Eqs.~(\ref{eqn:g-regular-reducible-irreducible-decomposition})
and~(\ref{eqn:pauli-principle-condition-local})
become
\begin{equation}
n_{\rm s}^2 = 
-\lim\limits_{\delta {\bf r} \to \boldsymbol{0}}
\left[
\chi_{\rm red, ss} (\delta {\bf r})
+
\chi_{\rm irr, ss} (\delta {\bf r})
\right],
\label{eqn:pauli-principle-condition-translationally-invariant}
\end{equation}
for the EoS that obeys the Pauli principle, and
\begin{equation}
g_{\rm s s^{\prime}}^{\rm  reg} ( \delta {\bf r})
= 
1 
+ 
\delta g_{\rm  red, s s^{\prime}}( \delta {\bf r} )
+
\delta g_{\rm  irr, s s^{\prime}}( \delta {\bf r} ),
\label{eqn:dens-dens-correl-transl-inv}
\end{equation}
for the normalized {\it spin}-resolved density-density correlation function. Here, $\delta {\bf r} = {\bf r}^\prime - {\bf r}$ labels the relative position and
$
\delta g_{\alpha,\rm s s^{\prime}}( \delta {\bf r} ) 
= 
\chi_{\alpha,\rm s s^\prime} (\delta {\bf r})/
n_{\rm s}   n_{\rm s^\prime} 
$
are the contributions from correlations.

Next, we discuss a translationally invariant continuum model for the BCS-BEC crossover and obtain explicitly the 
{\it spin}-resolved density-density response and the EoS.

\textit{BCS-BEC crossover:} We consider a {\it spin} mixture of fermions in $D$-dimensions with {\it spin} 
${\rm s} = \{\uparrow, \downarrow\}$ 
described by the Lagrangian density
\begin{eqnarray}
\mathcal{L} & = &\sum_{\rm s} \overline{\psi}_{\rm s}(x)\left[ \partial_\tau - \frac{\boldsymbol{\nabla}^2}{2m_{\rm s}} - \mu_{\rm s} \right] \psi_{\rm s}(x) \\
&-& g \overline{\psi}_{\uparrow}(x) \overline{\psi}_{\downarrow}(x)\psi_{\downarrow}(x) \psi_{\uparrow}(x) 
-\sum_{\rm s} j_{\rm s}(x)n_{\rm s}(x) \nonumber
\label{Eqn:Lagrangian_Density}
\end{eqnarray}
defined in the volume $L^D$.
Here, $\overline{\psi}_{\rm s} (x)$ and $\psi_{\rm s} (x)$, having dimensions $L^{-D/2}$,
are the Grassmann fields of {\it spin} $\rm s$ describing field operators $\hat\psi^\dagger_{\rm s} (x) $ and $\hat\psi_{\rm s} (x) $ at $x = ({\bf r}, \tau)$, with ${\bf r}$ as the position and $\tau \in [0,\beta]$ as the imaginary time.
Furthermore, $\mu_{\rm s}$ represents the chemical potential, $m_{\rm s}$ the mass of the {\it spin} species and $g\ge 0$ denotes the local $\rm SU(2)$-invariant attractive interaction strength with dimensions of energy times volume.
We use $\beta = 1/T$ as inverse temperature and units where $k_{\rm B} = 1 = \hbar$.

The Lagrangian density $\mathcal{L}$ includes a coupling between real 
{\it spin}-dependent source fields $j_{\rm s}(x)$
with dimensions of energy, 
and real density fields $n_{\rm s}(x)$ 
with dimensions of inverse volume. 
This leads to the dimensionless action $\mathcal{S}[\overline{\psi},\psi;j_{\rm s}] = \int {\rm d}x \mathcal{L}$ with $\int {\rm d}x = \int_0^\beta {\rm d}\tau \int_{L^D} {\rm d}^Dr$ used to obtain the source-dependent thermodynamic potential $\Omega [j_{\rm s}] = -T\ln \mathcal{Z}[j_{\rm s}]$, where the source-dependent partition function is 
$\mathcal{Z}[j_{\rm s}] = \oint \mathcal{D}[\overline{\psi},\psi]e^{-\mathcal{S}[\overline{\psi},\psi;j_{\rm s}]}$.

To compute the {\it spin}-resolved density-density correlator in the superfluid state,
we decouple the interaction in terms of the complex pair field $\Delta(x)$ via a Hubbard-Stratonovich transformation~\cite{Stratonovich-1957,Hubbard-1959}, 
and go to Fourier space. 
Integrating out the fermions and the pair
fluctuations, see End Matter, leads to the action contribution
\begin{equation}  
{\mathcal{S}}_{2} [ j_{\rm s} ] = \frac{\beta}{2} \sum_q {\bf j}^{\rm T}_{-q} \widetilde{\boldsymbol{\chi}}(q) {\bf j}_q,
\label{Eqn:GF-Action-Sources}
\end{equation}
where the {\it spin}-resolved density-density tensor, with dimensions of inverse energy, 
is 
\begin{equation}
\widetilde{\boldsymbol{\chi}}(q) = \widetilde{\boldsymbol{\chi}}_{0} (q) + \widetilde{\boldsymbol{\chi}}_{\rm irr}(q).
\label{eqn:spin-resolved-correlations}
\end{equation}
Here, ${\bf j}^{\rm T}_{q}$ denotes the Fourier transform of ${\bf j}^{\rm T} (x) = 
\left(j_{\uparrow} (x), j_{\downarrow} (x) \right)$, and $q = ({\bf q},iq_\ell)$, where 
${\bf q}$ is the momentum and $q_\ell$ represents the bosonic Matsubara frequency.
In Eq.~(\ref{eqn:spin-resolved-correlations}),
$\widetilde{\boldsymbol{\chi}}(q)$ is an approximate, Pauli-principle preserving and gauge-invariant solution to the
generalized Bethe-Salpeter equation~\cite{Bethe-1951}. 
The matrix elements of 
$\widetilde{\boldsymbol{\chi}}(q)$ are 
$
\widetilde{\chi}_{\rm ss^\prime}(q) = L^D \int_0^\beta {\rm d}\delta\tau \int {\rm d}^D\delta r \chi_{\rm ss^\prime}(\delta {\bf r}, \delta \tau) e^{-i\mathbf{q}\delta\mathbf{r} + iq_\ell \delta \tau},
$
describing the Fourier transform of the real space density-density correlators $\chi_{\rm ss^\prime}(\delta {\bf r}, \delta \tau)
= \langle \delta \hat n_{\rm s} 
(\boldsymbol{0},0) \delta \hat n_{\rm s^\prime} (\delta {\bf r}, \delta \tau) \rangle$, 
with dimensions of $L^{-2D}$.

The matrix $\widetilde{\boldsymbol{\chi}}_{0}(q)$ describes the connected reducible {\it spin}-resolved density-density response including
both $\chi_{\rm red, s s^\prime} (\delta {\bf r})$ from Eq.~(\ref{eqn:reducible-chi}) and a constant term arising 
from the singular part of $\chi_{\rm s s^\prime} (\delta {\bf r})$, defined in Eq.~(\ref{eqn:chi-regular-irregular-decomposition}),
when $n_{\rm s} ({\bf r}) = n_{\rm s}$.
The second term 
$\widetilde{\boldsymbol{\chi}}_{\rm irr}(q)$ 
represents the two-particle irreducible contribution 
describing collective excitations, many-body scattering states and vertex corrections, as discussed in Eq.~(\ref{eqn:irreducible-chi-real-space}) for its 
real-space counterpart. 

To obtain the Gaussian fluctuation (GF) EoS, arising only from pair fluctuations, we set $j_{\rm s} = 0$ and use 
$\Omega = - T \ln {\mathcal Z}[j_{\rm s} = 0]$ to give
$
n_{\rm s} =  
n_{\rm s, sp}
+ 
n_{\rm s, fl}.
$
Here, the first term $n_{\rm s, sp}$ comes from the saddle-point and the second 
$n_{\rm s, fl}$ is due to pair fluctuations, see 
Eq.~(\ref{eqn:GPF-EoS}) in the End Matter section.
The GF EoS is globally gauge-invariant, in the 
sense that it depends only on the order-parameter modulus 
$\vert \Delta_0 \vert$, but 
violates the Pauli principle, because it does not include the irreducible two-particle contributions.

Making the identifications 
$
\lim_{\delta {\bf r} \to \boldsymbol{0}} 
\chi_{\rm red, s s^{\prime}} (\delta{\bf r})
= \sum_q \widetilde\chi_{\rm red,ss}(q)/(\beta L^{2D})
$
for the reducible part 
and 
$\lim_{\delta {\bf r} \to \boldsymbol{0}} 
\chi_{\rm irr, s s^{\prime}} (\delta{\bf r})
= 
\sum_{q} \widetilde\chi_{\rm irr, ss}(q)/(\beta L^{2D}),
$ 
for the irreducible component, Eq.~(\ref{eqn:pauli-principle-condition-translationally-invariant}) becomes
\begin{equation}
 n_{\rm s}^2 = -\frac{1}{ \beta L^{2D}}\sum_q \widetilde\chi_{\rm red, ss}(q) - \frac{1}{\beta L^{2D}}\sum_{q} \widetilde\chi_{\rm irr, ss}(q),
 \label{eqn:equation-of-state-pauli-principle}
\end{equation}
representing the Pauli-respecting (PR) 
EoS, which is also gauge invariant.

We discussed general results for the {\it spin}-resolved density-density response in the BCS-BEC crossover for SU(2) invariant (s-wave) contact interactions at any spatial dimension, temperature, interaction, mass or population imbalance. 
\begin{table}[t!]
\centering
\begin{tabular}{|c||c|c|c|c|}
\hline
Case & Approximation & Gauge & Pauli & Line style \\ \hline
I  & SP EoS + only $\chi_{ss^\prime}^{(0)}$ & $\checkmark$ & $\checkmark$ & dashed black \\ \hline
II  & SP EoS + full $\chi_{ss^\prime}$ & $\checkmark$ & $\mathit{\times}$ & dotted green \\ \hline
III & GF EoS + full $\chi_{ss^\prime}$ & 
$\checkmark$ & $\times$ & dash-dotted red \\ \hline
IV & PR EoS + full $\chi_{ss^\prime}$ &  
$\checkmark$ & $\checkmark$ & solid blue \\ \hline
\end{tabular}
\caption{
Table for different cases covering distinct levels of approximation, depending on the preservation $(\checkmark)$ or violation $(\times)$ of gauge invariance (Gauge) and the Pauli principle (Pauli) of the approximation used to compute the correlation
function $\chi_{\rm ss^\prime}$.
Gauge invariance applied to EoS ($\chi_{\rm ss^\prime}$) is global (local).
The corresponding line styles used in the subsequent figures are
also indicated.
}
\label{tab:comparison-of-approaches}
\end{table}
Next, we investigate
$
\chi_{\rm s s^\prime} (\delta {\bf r}) = 
\langle \delta{\hat n}_{\rm s}({\bf 0}) 
\delta{\hat n}_{\rm s^\prime}({\delta \bf r}) \rangle
$
and $g_{\rm s s^\prime} (\delta {\bf r})$
for the BCS-BEC crossover in two dimensions (2D) particularized to $T=0$, equal masses and populations to compare with recent experiments~\cite{Yefsah-2025-3} close to this limiting case. 
The solutions for different EoS in 2D are discussed in Fig.~\ref{fig:Equation_of_State} in the End Matter section.

\textit{Correlation functions in 2D:} 
Using rotational invariance at $T = 0$, we find
\begin{equation}
\chi_{s s^\prime} (\vert \delta {\bf r} \vert) = 
\frac{1}{L^2} \int_{-\infty}^\infty \frac{{\rm d}q_\ell}{2\pi}\int_0^\infty \frac{{\rm d}Q}{2\pi} \hspace{.1cm}Q \widetilde{\chi}_{\rm ss^\prime}(Q,iq_\ell) J_0(Q\vert\delta \mathbf{r}\vert),
\end{equation}
where integration over the azimuthal angle leads to the the 
zeroth order Bessel function $J_0 (Q \vert \delta {\bf r}\vert )$ with $\vert \delta {\bf r} \vert$ and $Q = \vert {\bf q} \vert$ being the modulus of $\delta {\bf r}$ and ${\bf q}$, respectively.

We apply the Lippmann-Schwinger relation~\cite{SaDeMelo-2006} to replace the interaction 
$g$ by the two-body bound state energy $\varepsilon_{\rm B} > 0$, and use the Fermi momentum (energy) $k_{\rm F}$ ($\varepsilon_{\rm F}$) to relate $\varepsilon_{\rm B}$ to the 2D scattering length $a$ as 
$\ln k_{\rm F}a = \ln(8\varepsilon_{\rm F}/\varepsilon_{\rm B})/2 - \gamma_{\rm E}$, with
$\gamma_{\rm E}$ as Euler constant.

\begin{figure}[t!]
\centering
\includegraphics[width=1\linewidth]{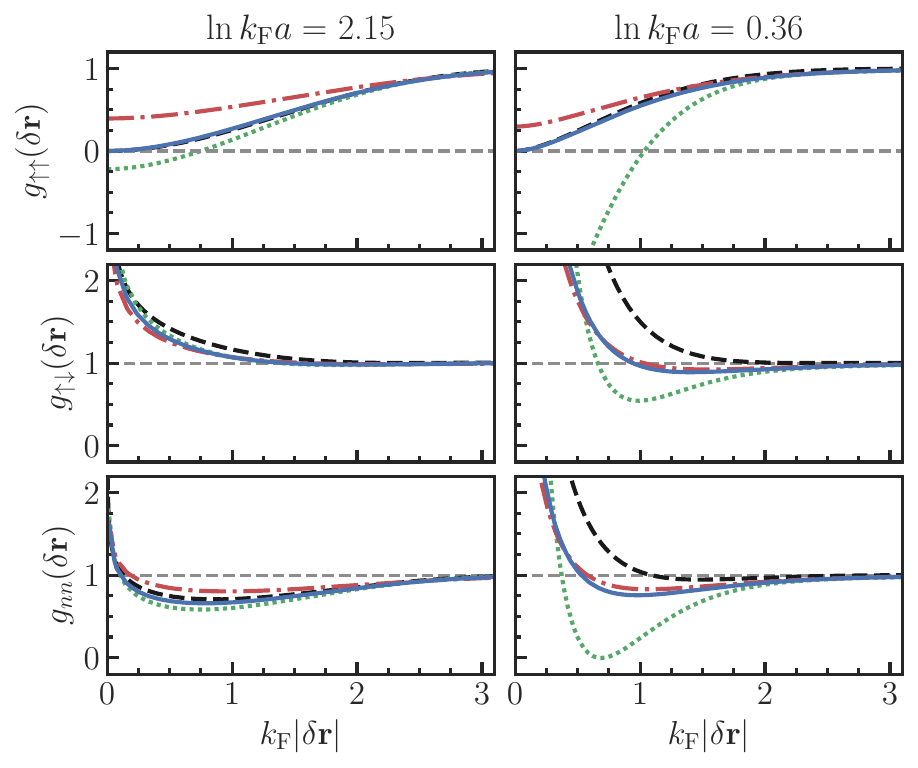}
\caption{
Plots of spatial behavior of the correlation functions
$g_{\rm s s^\prime} (\delta {\bf r})$ and of 
$g_{nn} (\delta {\bf r})
$ 
for translationally and rotationally invariant
systems. 
The interaction parameters are $\ln k_{\rm F} a = 2.15$ 
(BCS regime) and
$\ln k_{\rm F} a = 0.36$ (crossover region).
The line types indicate the level of approximation used,
see Table~\ref{tab:comparison-of-approaches}.
}
\label{fig:correlation-functions-comparison-of-approximations}
\end{figure}

In Fig.~\ref{fig:correlation-functions-comparison-of-approximations}, we illustrate the behavior of 
$g_{\rm s s^\prime} (\delta {\bf r})$, defined in Eq.~(\ref{eqn:dens-dens-correl-transl-inv}),
and of the full density-density correlation function $g_{nn} (\delta {\bf r}) = \sum_{\rm s s^\prime} g_{\rm s s^\prime} (\delta {\bf r})$
versus separation $k_{\rm F} \vert \delta {\bf r} \vert$.
The interaction parameters are $\ln k_{\rm F} a = 2.15$ 
(BCS regime) and $\ln k_{\rm F} a = 0.36$ (crossover region).
The different level of approximations and line types are summarized in 
Table~\ref{tab:comparison-of-approaches}.
For $g_{\uparrow\uparrow} (\delta {\bf r})$, the dotted green lines (case II) and dash-dotted red lines (case III) violate the Pauli exclusion principle, while the dashed black lines (case I) and the solid blue lines (case IV) satisfy it. For 
$g_{\uparrow\downarrow} (\delta {\bf r})$ 
the dotted green lines (case II) and the 
dash-dotted red lines (case III) 
do not differ substantially from the dashed black lines (case I) and the solid blue lines (case IV) in the BCS regime
$(\ln k_{\rm F} a = 2.15)$, but start to behave differently in the crossover
region $(\ln k_{\rm F} a = 0.36)$ and beyond, due to contributions from 
the two-particle irreducible terms.
The main purpose of Fig.~\ref{fig:correlation-functions-comparison-of-approximations} is to show that the approximations used in cases
II and III of Table~\ref{tab:comparison-of-approaches}
are gauge invariant but violate the Pauli principle, while the approaches in cases I and IV satisfy both gauge invariance and the Pauli principle.

\begin{figure}[t!]
\centering
\includegraphics[width=1\linewidth]{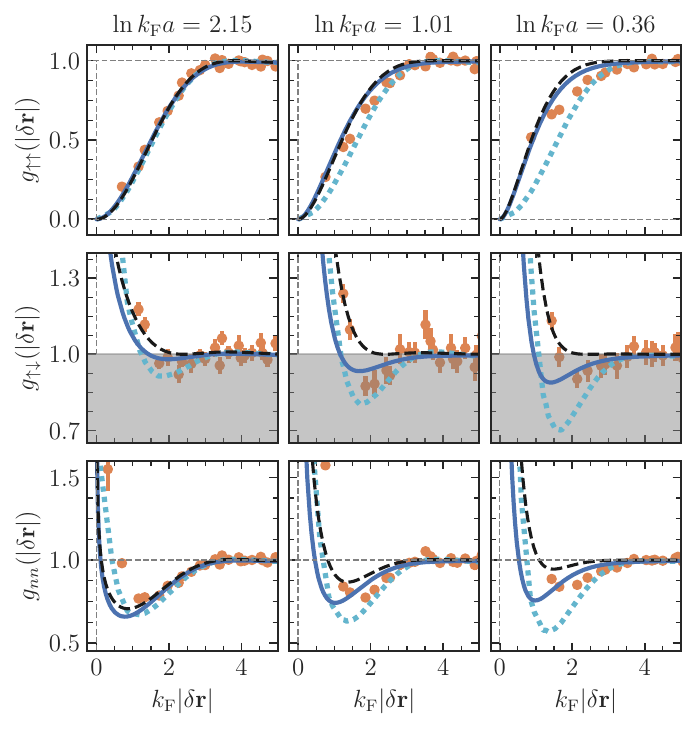}
\caption{Numerical results for 
$g_{\uparrow\uparrow} (\delta {\bf r})$ ranging from 
$0$ and $1$ (top panel),
$g_{\uparrow\downarrow} (\delta {\bf r})$ 
(middle panel) ranging from $0.7$ to $1.4$
and 
$g_{nn} (\delta {\bf r})$ (bottom panel) ranging
from $0.5$ to $1.5$ versus distance 
$k_{\rm F} \vert \delta {\bf r}\vert$ ranging from 
$0$ to $5$.
The dashed black (solid blue) lines illustrate case ${\rm I}$
(IV) in Table~\ref{tab:comparison-of-approaches}.
The gray region in the middle panel shows that 
$g_{\uparrow\downarrow} (\delta {\bf r}) \ge 1$ with no minimum 
for the dashed black line, while 
$g_{\uparrow\downarrow} (\delta {\bf r})$ has a clear
minimum below 1 for the solid blue line.
The dotted cyan line and the solid oranges circles describe Monte Carlo and experimental data from Ref.~\cite{Yefsah-2025-3,Yefsah-Data-2026} near $T/T_F = \{0.11, 0.09, 0.08 \}$ for 
$\ln k_{\rm F} a = \{ 2.15, 1.01, 0.36 \}$. 
}
\label{fig:correlation-functions-pauli-preserving}
\end{figure}
In Fig.~{\ref{fig:correlation-functions-pauli-preserving}}, we depict only cases I and IV,
showing the correlation functions
$g_{\uparrow\uparrow} (\vert \delta {\bf r} \vert)$,
$g_{\uparrow\downarrow} (\vert \delta {\bf r} \vert)$,
$g_{nn} (\vert \delta {\bf r} \vert)$
versus $k_{\rm F} \vert \delta {\bf r} \vert$ for scattering parameters
$\ln {k_{\rm F} a} = \{2.15, 1.01, 0.36\}$,
and a comparison to Monte Carlo (dotted cyan
lines) and experimental (solid orange circles) 
data from Ref.~\cite{Yefsah-2025-3,Yefsah-Data-2026} is made. An extended version of Fig.~\ref{fig:correlation-functions-pauli-preserving}
is seen in Fig.~\ref{fig:correlation-functions-pauli-preserving_EM},
see End Matter section.

In Fig.~\ref{fig:correlation-functions-pauli-preserving} (top panel),
we demonstrate that the difference between cases I and
IV is small in $g_{\uparrow\uparrow} (\delta {\bf r})$ for the parameters shown.
Due to the absence of triplet pairing in our model, that is, $F_{\uparrow\uparrow}(\delta\mathbf{r}) = 0$, 
the only contributions to 
$g_{\uparrow\uparrow} (\delta {\bf r})$, for case I,
come from the two-particle reducible correlations 
$\delta g_{\rm red, \uparrow \uparrow} (\delta {\bf r})
= \chi_{{\rm red}, \uparrow \uparrow} (\delta {\bf r}) /n_\uparrow^2 = 
- \vert G_{\uparrow \uparrow} (\delta {\bf r}) \vert^2/n_\uparrow^2$.
This term is always comparable to the sum of 
$\delta g_{\rm red, \uparrow \uparrow} (\delta{\bf r})$ and 
$
\delta g_{\rm irr, \uparrow \uparrow} (\delta{\bf r}) = 
\chi_{{\rm irr},\uparrow \uparrow} (\delta {\bf r}) /n_\uparrow^2 
$ for case IV in the parameter range shown.
The spatial extent $k_{\rm F} r_{\rm p}$ 
describes the size of 
{\it Pauli hole} when $g_{\uparrow \uparrow} (\delta {\bf r}) = 0.5$. 

In Fig.~\ref{fig:correlation-functions-pauli-preserving}
(middle panel), we show that differences
between cases I and IV are more dramatic in $g_{\uparrow\downarrow} (\delta {\bf r})$
for the parameters shown.
In case ${\rm  I}$, the only contribution 
to $g_{\uparrow\downarrow} (\delta {\bf r})$
comes from the two-particle reducible correlations
$\delta g_{\rm red,\uparrow \downarrow} = 
\chi_{{\rm red}, \uparrow \downarrow} (\delta {\bf r}) /n_\uparrow n_{\downarrow}
=
\vert F_{\uparrow \downarrow} (\delta {\bf r}) \vert^2/n_{\uparrow} n_{\downarrow}
$
being always positive.
There is no negative contribution arising, since $G_{\uparrow\downarrow}(\delta\mathbf{r})=0$ due to 
{\it spin}-projection conservation along the {\it spin} $z$ axis.
However, in case ${\rm IV}$,
there are two contributions: $\delta g_{\rm red,\uparrow \downarrow} (\delta {\bf r})$, 
which is always positive, 
and $\delta g_{\rm irr, \uparrow \downarrow} (\delta {\bf r})$ which is always negative (anti-bunching) 
due to the coupling to pair fluctuations.
Case ${\rm I}$ (dashed black line) always gives $g_{\uparrow \downarrow} (\delta {\bf r}) \ge 1$ and does not produce a minimum in $g_{\uparrow \downarrow} (\delta {\bf r})$. 
The minimum in $g_{\uparrow\downarrow} (\delta{\bf r})$ arises 
from the interplay between two-particle reducible and irreducible contributions, where the latter are included only in case ${\rm IV}$ (solid blue line). 

\begin{figure}[t!]
\centering
\includegraphics[width=1\linewidth]{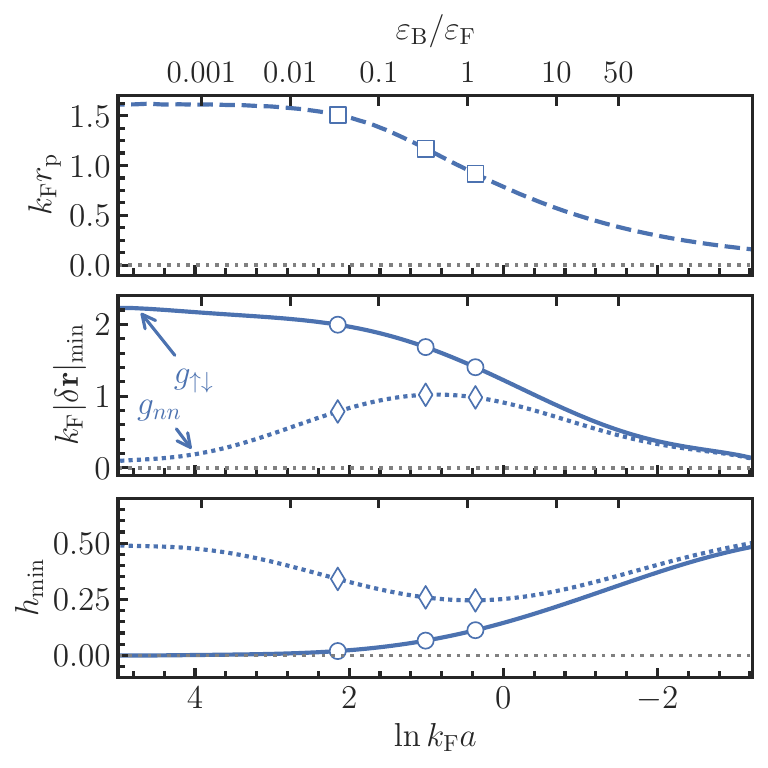}
\caption{Plots of $k_{\rm F} r_{\rm p}$
(top panel), $k_{\rm F} \vert \delta {\bf r} \vert_{\rm min}$
(middle panel), $h_{\rm min}$ (bottom panel), versus $\ln k_{\rm F} a$. The depth of the minimum is 
$h_{\rm min} = \vert g_{\uparrow\downarrow} 
(\delta{\bf r}_{\rm min}) - 1 \vert$
or 
$h_{\rm min} = \vert g_{nn} 
(\delta{\bf r}_{\rm min}) - 1 \vert$.
The solid blue lines refer to $g_{\uparrow\downarrow} (\delta {\bf r})$, the dotted blue lines represent 
$g_{\uparrow\downarrow} (\delta {\bf r})$, and
the dashed blue line reflects $g_{\uparrow\uparrow} (\delta {\bf r})$.
The blue circles, diamonds and squares describe the scattering
parameters used in Fig.~\ref{fig:correlation-functions-pauli-preserving}.
}
\label{fig:Minima}
\end{figure}

In Fig.~\ref{fig:correlation-functions-pauli-preserving}
(bottom panel), we establish that cases I and IV also differ substantially
for $g_{nn} (\delta{\bf r})$, for the parameters used, due to the differences
in $g_{\uparrow\downarrow} (\delta {\bf r})$.
In the range of parameters shown, both cases I and IV produce a minimum for $g_{nn} (\delta{\bf r})$
because of the increase in $g_{\uparrow\uparrow} (\delta{\bf r}) = g_{\downarrow\downarrow} (\delta{\bf r})$ and the decrease
in $g_{\uparrow\downarrow} (\delta{\bf r}) = 
g_{\downarrow\uparrow} (\delta{\bf r})$ with growing 
$k_{\rm F} \vert \delta {\bf r} \vert$ from $0$ to $\lesssim 1$. However, 
the location and depth of the minimum in $g_{nn} (\delta{\bf r})$
is different from that of $g_{\uparrow\downarrow} (\delta{\bf r})$.
To clarify this behavior further, 
we discuss next the location and depth of 
the minima in $g_{\uparrow \downarrow} (\delta {\bf r})$
and $g_{n n} (\delta {\bf r})$, as well as 
$k_{\rm F}r_{\rm p}$ characterizing the {\it Pauli hole} in $g_{\uparrow \uparrow} (\delta {\bf r})$.

Although a comparison of our work at $T =0$ with current Monte Carlo (MC) and experimental results for $T/T_F \approx 0.1$ from Ref.~\cite{Yefsah-2025-3} is good in the BCS regime $(\ln k_{\rm F} a = 2.15)$, deviations occur as the crossover region $(\ln k_{\rm F} a = 0.36)$ is approached. These differences are attributed to finite temperature effects and to experiments no longer being in 2D as the crossover region is entered. 
Furthermore, MC calculations potentially deal
with a different equation of state that exaggerates the minimum in 
$g_{nn} (\delta \vert {\bf r} \vert )$. See further comparison in Fig.~\ref{fig:correlation-functions-pauli-preserving_EM} in the End Matter.

In Fig.~\ref{fig:Minima}, we show $k_{\rm F} r_{\rm p}$ (top panel), the position of the minimum
$k_{\rm F} \vert \delta{\bf r}\vert_{\rm min}$ (middle panel)
in $g_{\uparrow\downarrow} (\delta{\bf r})$ and 
$g_{nn} (\delta{\bf r})$, as well as, its depth $h_{\rm min}$ (bottom panel) versus $\ln k_{\rm F} a$,
for case ${\rm IV}$ only. 
As seen in the top panel, $k_{\rm F} r_{\rm p}$ becomes smaller with growing $\ln k_{\rm F} a$ approaching zero asymptotically in the extreme Bose regime for low densities.
The location and depth of the minimum in 
$g_{\uparrow\downarrow} (\delta{\bf r})$ and $g_{nn} 
(\delta{\bf r})$ differ substantially in the BCS regime, 
because $k_{\rm F} r_{\rm p}$ is sufficiently large to 
produce a deeper minimum located at shorter distances 
in $g_{nn} (\delta {\bf r})$. However, as $k_{\rm F} r_{\rm p}$
decreases with increasing $\ln k_{\rm F} a$, 
the location $k_F \vert \delta{\bf r}\vert_{\rm min}$ 
and depth $h_{\rm min}$ of the minimum in $g_{nn} (\delta{\bf r})$ 
approaches those of $g_{\uparrow\downarrow} (\delta{\bf r})$, since
the latter correlations become more important. 

\textit{Conclusions:}
We developed a general theory of 
{\it spin}-dependent density-density correlations, 
that is valid for any temperature, interactions, dimensions and mass or population status
of Fermi {\it spin}-mixtures.  
We used gauge invariance and the Pauli principle to 
establish constraints on the {\it spin}-dependent density-density correlations that are consistent with the fluctuation-dissipation and Wick's theorem.
As an example, we studied
the {\it spin}-dependent density-density correlations from the BCS to the Bose regime in two dimensions at zero temperature, inspired by experiments in 
$^6{\rm Li}$~\cite{Yefsah-2025-3}. We showed that two-particle irreducible contributions, such as, collective excitations, many-particle scattering, and vertex corrections are responsible for
the differences between saddle-point 
and fluctuation approaches.
Two-particle irreducible terms slow down the 
closing of the {\it Pauli hole}
with increasing interactions when compared to 
saddle-point results; they are also responsible for the minimum in the opposite-{\it spin} density-density correlations, which is absent for saddle-point theories, in agreement with recent
experiments~\cite{Yefsah-2025-3}.

As an outlook, we will perform calculations at finite temperatures, which are technically more challenging, because it is necessary to handle branch cuts and poles (collective excitations) of the correlation function with the additional complication of including vortices and antivortices in two-dimensions.

{\it Note added:} During the writing of our work, we became aware of a very recent pre-print~\cite{Castin-2026}, that also addresses experiments~\cite{Yefsah-2025-3} using a different approach.

\textit{Acknowledgments:} We thank Joshua Krauss, Flavia Braga Ramos and Sejung Yong for discussions.
C. A. R. SdM thanks the German Research Foundation (Mercator Fellowship) for support.
We acknowledge financial support by the Deutsche Forschungsgemeinschaft (DFG, German Research Foundation) via the Collaborative Research Center SFB/TR185 
(Project No.\ 277625399).

\bibliography{References}

\begin{widetext}
    \vspace{.01cm}
\end{widetext}

\mbox{~}
\clearpage

\appendix
\section*{End Matter}
\renewcommand{\theequation}{A\arabic{equation}}
\setcounter{equation}{0}

\textit{Response functions:}
Performing a Fourier transformation on the fermions
\begin{equation}
c_{\rm s} ({\mathbf{k},ik_n}) = \int_0^\beta \frac{{\rm d}\tau}{\beta} \int \frac{{\rm d}^Dx}{\sqrt{L^D}} \psi_{\rm s} (\mathbf{x},\tau) e^{-i {\bf k} \cdot \mathbf{x} + ik_n \tau },
\end{equation}
and integrating them through a standard procedure~\cite{Engelbrecht-1993}, leads
to the dimensionless effective action 
\begin{equation}
S_{\rm eff} [j_{\rm s}] = \beta L^D \sum_{q} \frac{|\Delta_q|^2}{g} - \sum_{kk^\prime} \ln{\rm det} \left[ -\beta \mathbf{G}^{-1} \right].
\label{Eqn:BCS-Action}
\end{equation}
Here, the label $q = ({\bf q},iq_\ell)$
represents momentum ${\bf q}$ and 
bosonic Matsubara frequency $q_\ell$, while 
``$\rm det$" is determinant over {\it spin} indices.
Using the momentum space relation 
$
\Delta_q = \int_0^\beta \frac{{\rm d}\tau}{\beta}\int \frac{{\rm d}^Dx}{L^D} \Delta(x)e^{-i\mathbf{q}\cdot\mathbf{x} + iq_\ell \tau},
$
we obtain
\begin{equation}
\mathbf{G}^{-1}(k,k^\prime) = 
\begin{pmatrix}
\alpha_{\uparrow} (k, k^\prime)
- j_{\uparrow,k-k^\prime} & \Delta_{k-k^\prime} \\ \overline{\Delta}_{k^\prime-k} & \alpha_{\downarrow}(k,k^\prime) + j_{\downarrow,k-k^\prime}
\end{pmatrix},
\end{equation}
with diagonal terms
$\alpha_{\uparrow} (k, k^\prime)
= 
(ik_n - \xi_{\uparrow,{\bf k}})\delta_{k,k^\prime}$ 
and $\alpha_{\downarrow}(k,k^\prime) = (ik_n + \xi_{\downarrow, {\bf k}})\delta_{k,k^\prime}$
involving the kinetic energies $\xi_{{\rm s}, {\bf k}} = {\bf k}^2/2m_{\rm s} - \mu_{\rm s}$ with respect to chemical potentials $\mu_{\rm s}$. We use the notation $k = ({\bf k},ik_n)$ for 
momentum ${\bf k}$ and 
fermionic Matsubara frequency $k_n$.
The term $j_{{\rm s},q}$ is related to $j_{\rm s} (x)$
via the transformation
$
j_{{\rm s},\mathbf{k},ik_n} = \int_0^\beta \frac{{\rm d}\tau}{\beta} \int \frac{{\rm d}^Dx}{L^D} j_{\rm s} (\mathbf{x},\tau) e^{-i {\bf k} \cdot \mathbf{x} + ik_n \tau }.
$

Since our goal is to compute the locally gauge-invariant 
{\it spin}-resolved density-density correlation functions, 
it is essential to include the gauge freedom of the pair
field. We decompose $\Delta_q = \Delta_0\delta_{q,0} + \lambda_q + i\theta_q$, where $\Delta_0$ is the complex time-independent and spatially-uniform order parameter and $\lambda_q$
and $\theta_q$ are pair-fluctuation real fields necessary to guarantee the local 
gauge invariance of the full {\it spin}-dependent density-density correlations.

We expand $S_{\rm eff} [j_{\rm s}]$ to quadratic order in $\lambda_q$ and
$\theta_q$, followed by an expansion to Gaussian order in $j_{{\rm s},q}$. This procedure gives three contributions
\begin{equation}
\mathcal{S}_{\rm eff} [j_{\rm s}] = 
\mathcal{S}_0 + \mathcal{S}_1 [j_{\rm s}] 
+ \mathcal{S}_2 [j_{\rm s}, \lambda, \theta] 
+ \mathcal{O}(j_{\rm s}^3,\lambda^3,\theta^3).
\end{equation}
The leading order term of the effective action is
\begin{equation}
\mathcal{S}_{0} = \beta L^D \frac{|\Delta_0|^2}{g} - \sum_{k} \ln\det \left[ -\beta \mathbf{G}^{-1}_{0}(k) \right].
\end{equation}
where the saddle-point inverse Green's function is
\begin{equation}
\mathbf{G}^{-1}_{0}(k) = 
\begin{pmatrix}
\alpha_{\uparrow} (k,k) & \Delta_{0} \\ \overline{\Delta}_{0} & \alpha_{\downarrow}(k,k) 
\end{pmatrix}.
\end{equation}
Extremizing $\mathcal{S}_0$, that is, $\partial \mathcal{S}_0/\partial {\overline \Delta}_0 = 0$, 
leads to
\begin{equation}
\Delta_0 = \frac{g}{\beta L^D} \sum_{k} {\rm tr}\left\{ \mathbf{G}_0(k) \frac{\partial \mathbf{G}_0^{-1}(k)}{\partial \overline{\Delta}_0} \right\}.
\label{eqn:Order-parameter-equation}
\end{equation}
The first order correction is 
\begin{equation}
\mathcal{S}_{1} [j_{\rm s}] = - \sum_{k, \rm s} G_{0,\rm ss}(k) j_{{\rm s},0},
\label{eqn:Linear-action-contribution}
\end{equation}
where $G_{0, \rm ss} (k)$ is a diagonal matrix element of 
${\bf G}_0 (k)$. The linear terms in $\lambda$ and $\theta$ vanish due to
the saddle-point condition in Eq.~(\ref{eqn:Order-parameter-equation}).
The second term is 
\begin{equation}\label{eqn:effective-action-2}
\mathcal{S}_{2} [{j_{\rm s}}, \lambda, \theta] = \frac{\beta}{2} \sum_q \boldsymbol{\phi}^{\rm T}_{-q} \boldsymbol{\Pi}(q)\boldsymbol{\phi}_q,
\end{equation}
where the fields
$
\boldsymbol{\phi}_{-q}^{\rm T} = \begin{pmatrix} \boldsymbol{j}^{\rm T}_{-q} & \boldsymbol{\eta}^{\rm T}_{-q} \end{pmatrix}
$
have dimensions of energy and
include the real and imaginary fluctuations of the pair field
$\boldsymbol{\eta}^{\rm T}_{-q} = \begin{pmatrix} \lambda_{-q} & \theta_{-q} \end{pmatrix}$; while the source currents
$\boldsymbol{j}^{\rm T}_{-q} = \begin{pmatrix} j_{\uparrow,-q} & j_{\downarrow,-q} \end{pmatrix}$ play the role of
a non-uniform {\it fluctuation} in the chemical potentials
$\mu_{\uparrow}$ and $\mu_{\downarrow}$.
\begin{figure}[t!]
\centering
\includegraphics[width=1\linewidth]{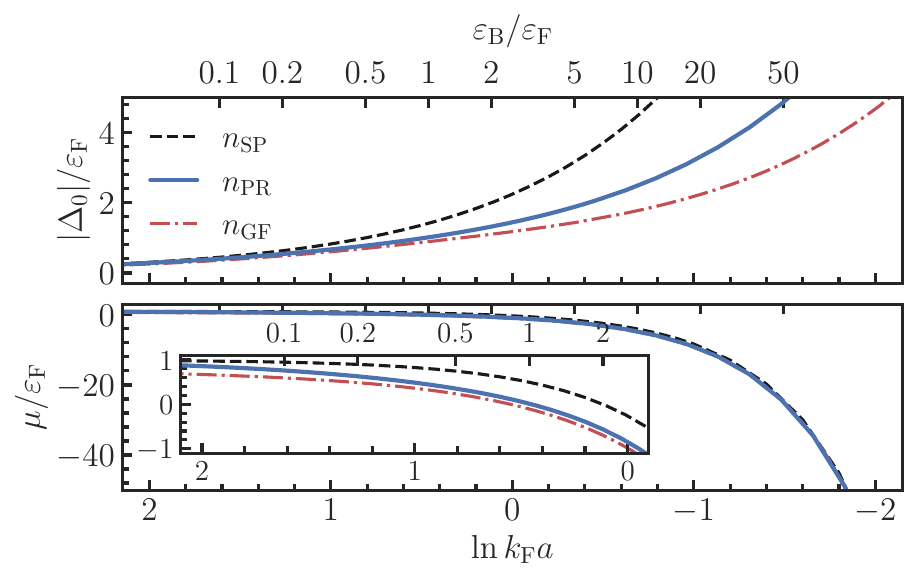}
\caption{Self-consistent results for $|\Delta_0|/\varepsilon_{\rm F}$ (top panel) and $\mu/\varepsilon_{\rm F}$ (bottom panel) versus $\ln k_{\rm F}a$ (lower $x$-axis) or binding energy $\varepsilon_{\rm B}/\varepsilon_{\rm F}$ (upper $x$-axis).
The dashed black line shows the saddle-point (SP), the dash-dotted red line describes the Gaussian fluctuations (GF), while the solid blue line represents the Pauli-respecting (PR) equation of state (EoS).}
\label{fig:Equation_of_State}
\end{figure}
\begin{figure*}[ht!]
\centering
\includegraphics[width=1\linewidth]{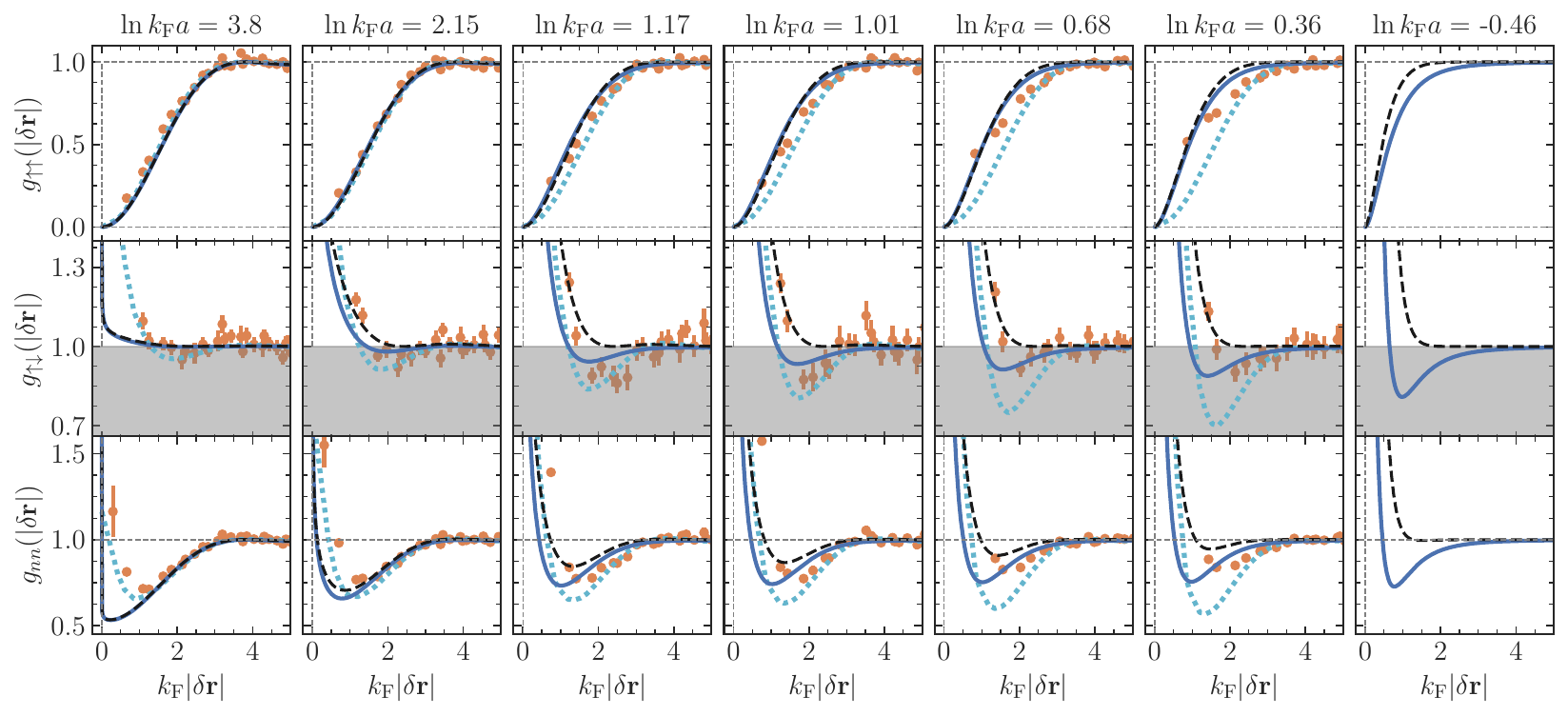}
\caption{Numerical results for 
$g_{\uparrow\uparrow} (\delta {\bf r})$ 
(top panel),
$g_{\uparrow\downarrow} (\delta {\bf r})$ 
(middle panel)
and 
$g_{nn} (\delta {\bf r})$ (bottom panel) 
versus
$k_F \vert \delta {\bf r}\vert$ with
$\ln k_{\rm F} a$ ranging
from 
$3.01$ (BCS regime) to $-0.46$ (BEC region). 
Line types and points have same meaning as in Fig.~\ref{fig:correlation-functions-pauli-preserving}.}
\label{fig:correlation-functions-pauli-preserving_EM}
\end{figure*}
The elements of ${\boldsymbol{\Pi}}$, 
having dimensions of inverse energy, are
\begin{equation}
\Pi_{ab}(q) = 
v_{ab}
+ \frac{1}{\beta} \sum_k {\rm tr}\left\{ \mathbf{G}_0(k)\boldsymbol{\gamma}_a\mathbf{G}_0(k+q)\boldsymbol{\gamma}_b \right\},
\end{equation}
where $v_{ab} = 2 \delta_{ab}(\delta_{ax} + \delta_{ay})/(L^D g)$, with
$\{ a, b \} \in \{\uparrow, \downarrow, x, y \}$.
The $\boldsymbol{\gamma}_a$ are $2\times 2$ matrices defined as
$\boldsymbol{\gamma}_{\uparrow} = -(\mathbf{I} + \boldsymbol{\sigma}_z)  /2$, $\boldsymbol{\gamma}_{\downarrow} = (\mathbf{I} - \boldsymbol{\sigma}_z)/2$,
$\boldsymbol{\gamma}_x = \boldsymbol{\sigma}_x$, $\boldsymbol{\gamma}_y = \boldsymbol{\sigma}_y$,
where ${\bf I}$ is the identity, and $\boldsymbol{\sigma}_x$, 
$\boldsymbol{\sigma}_y$ and $\boldsymbol{\sigma}_z$ are the 
Pauli matrices.

The notation above simplifies the description of ${\boldsymbol{\Pi}}$, but hides its physical meaning, 
thus we write
\begin{eqnarray}
\boldsymbol{\Pi}(q) && = \left(\begin{array}{cc|cc}
\Pi_{\uparrow\uparrow}(q) & \Pi_{\uparrow\downarrow}(q) & \Pi_{\uparrow \rm x}(q) & \Pi_{\uparrow \rm y}(q) \\
\Pi_{\downarrow\uparrow}(q) & \Pi_{\downarrow\downarrow}(q) & \Pi_{\downarrow \rm x}(q) & \Pi_{\downarrow \rm y}(q) \\ \hline
\Pi_{\rm x\uparrow}(q) & \Pi_{\rm x\downarrow}(q) & \Pi_{ \rm xx}(q) & \Pi_{\rm xy}(q) \\
\Pi_{\rm y\uparrow}(q) & \Pi_{\rm y\downarrow}(q) & \Pi_{\rm yx}(q) & \Pi_{\rm yy}(q)
\end{array}\right) \\
&& = \left(\begin{array}{c|c}
\widetilde{\boldsymbol{\chi}}_{0}(q) & \boldsymbol{\Lambda}^{\rm T}(q) \\ \hline
\boldsymbol{\Lambda}(q) & \boldsymbol{\Gamma}(q)
\end{array}\right) \label{eqn:chi-zero}
\end{eqnarray}
in a block-diagonal form.
The matrix  elements $\boldsymbol{\widetilde\chi}_{0}(q)$ are simply source-source (jj) correlation functions, that is, the symbols $\{\uparrow,
\downarrow\}$ represent $\{ j_\uparrow, j_\downarrow \}$,
respectively.
The matrix $\boldsymbol{\Lambda} (q)$ describes the
coupling between the source terms $\{j_{\uparrow}, j_\downarrow\} \leftrightarrow\{\uparrow, \downarrow \}$ and the pair fluctuations 
$\{\lambda, \theta\} \leftrightarrow \{x, y\}$.
The block matrix $\boldsymbol{\Gamma} (q)$ 
represents the 
pair fluctuations only $\{\lambda, \theta\} \leftrightarrow \{x, y\}$.
Integrating out the fluctuation fields 
$\boldsymbol{\eta}_q$ leads to
\begin{equation}
\mathcal{S_{\rm eff}}[j_{\rm s}] = 
{\mathcal{S}}_{\rm GF} 
+ {\mathcal{S}}_{1} [j_{\rm s}] 
+ {\mathcal{S}}_{2} [j_{\rm s}],
\end{equation}
where the Gaussian fluctuation (GF) action is 
\begin{equation}
{\mathcal{S}}_{\rm GF} = \mathcal{S}_0 + \sum_q \ln\det\left[\beta\boldsymbol{\Gamma}(q)\right].
\label{eqn:action-gaussian-fluctuation}
\end{equation}
Here $\mathcal{S}_1$ is defined in Eq.~(\ref{eqn:Linear-action-contribution})
and $\mathcal{S}_{2} [j_{\rm s}]$ is given  Eq.~(\ref{Eqn:GF-Action-Sources}).

From $\mathcal{S}_{2} [j_{\rm s}]$, we can read off
$
\widetilde{\boldsymbol{\chi}}(q) = \widetilde{\boldsymbol{\chi}}_{0} (q) + \widetilde{\boldsymbol{\chi}}_{\rm irr}(q),
$
displayed in Eq.~(\ref{eqn:spin-resolved-correlations}) of the main text. The first term 
$\widetilde{\boldsymbol{\chi}}_{0} (q)$ contains the connected two-particle-reducible contributions and a constant term from the singular part of $\chi_{\rm s s^\prime} (\delta {\bf r})$. 
The second term
$\widetilde{\boldsymbol{\chi}}_{\rm irr}(q) = -\boldsymbol{\Lambda}^{\rm T}(-q)\boldsymbol{\Gamma}^{-1}(q)\boldsymbol{\Lambda}(q)$ 
represents the two-particle irreducible contribution describing collective excitations, many-body scattering states and vertex corrections arising from the coupling between density and pair fluctuations.
For instance, the integer zeros of ${\bf \Gamma} (q)$ represent the collective excitations due to pair fluctuations~\cite{SaDeMelo-2023}.
When relating $\widetilde{\chi}_{\rm s s^\prime} (q)$ to the dynamical structure factor tensor $S_{\rm s s^\prime} (q)$, the contribution
from $\chi_{{\rm irr},\rm s s^\prime} (q)$ is essential to fulfill the compressibility and f-sum rules~\cite{SaDeMelo-2011, He-2016}.

The action ${\mathcal S}_{\rm GF}$ in Eq.~(\ref{eqn:action-gaussian-fluctuation})
gives the GF equation of state (EoS)
\begin{equation}
n_{\rm s} =  \frac{1}{L^D\beta} \sum_k G_{0, \rm ss}(k)
-\frac{1}{2L^D\beta} \sum_q \frac{\partial}{\partial \mu_{\rm s}} 
\ln {\rm det} \left[ \beta {\boldsymbol \Gamma} (q) \right],
\label{eqn:GPF-EoS}
\end{equation}
where the first term comes from $\mathcal {S}_0$ and represents the saddle-point (SP) EoS, whereas
and the second is due to pair fluctuations.

\textit{Crossover in 2D:}
Two-dimensional ($D =  2$) systems exhibit a two-body bound state with binding energy 
$\varepsilon_{\rm B}$ for arbitrarily small contact interactions. In contrast, for three dimensions $(D = 3)$, there is a threshold for the emergence to two-body bound states.  Furthermore, in 3D, the inclusion of longitudinal fluctuations in $\lambda,\theta$ are sufficient to describe the BCS-BEC crossover at finite temperatures. However, in 2D, we must also include transverse fluctuations (vortices and antivortices) due to Berezinskii-Kosterlitz-Thouless (BKT) mechanism.
Thus, for now, we only investigate the experimentally relevant example of a 2D Fermi system with balanced masses $(m_\uparrow = m_\downarrow = m)$ and populations $(n_\uparrow = n_\downarrow = n/2)$ at $T=0$.

In Fig.~\ref{fig:Equation_of_State}, using
the Fermi energy $\varepsilon_{\rm F}$, we show self-consistent solutions of 
$\vert \Delta_0 \vert/\varepsilon_{\rm F}$ 
and $\mu/\varepsilon_{\rm F}$ for: saddle-point (SP) (dashed-black line), saddle-point plus Gaussian fluctuations (GF) (dashed-dotted red line) and Pauli-respecting (PR) (solid blue). 
The results for $\vert \Delta_0 \vert/\varepsilon_{\rm F}$ $(\mu/\varepsilon_{\rm F})$ are illustrated in the top (bottom) panel versus $\ln k_{\rm F}a$ ($\varepsilon_{\rm B}/\varepsilon_{\rm F})$ depicted as lower (upper) $x$-axis. 
The values of $\vert \Delta_0 \vert/\varepsilon_{\rm F}$ and $\mu/\varepsilon_{\rm F}$
for the PR equation of state (EoS) always lie between 
SP and GF results.

In Fig.~\ref{fig:correlation-functions-pauli-preserving_EM}, we present an extension of the data found in Fig.~\ref{fig:correlation-functions-pauli-preserving} of the main text. The interaction parameters range now from $\ln k_{\rm F}a = 3.01$ (deeper into the BCS region) to 
$\ln k_{\rm F} a = -0.46$ (further into the BEC regime). The plots in Fig.~\ref{fig:correlation-functions-pauli-preserving_EM} further support 
the findings in the main text.


\end{document}